\documentstyle[epsfig]{article}
\def\circa#1{\,\raise.3ex\hbox{$#1$\kern-.75em\lower1ex\hbox{$\sim$}}\,}

\newcommand{\Mnaif}{{\cal M}}
\newcommand{\Inaif}{{\cal I}}
\newcommand{\Hnaif}{{\cal H}}
\newcommand  \f  \varphi
\newcommand \bra {\langle}
\newcommand \ket {\rangle}
\newcommand{\be}{\begin{equation}}
\newcommand{\ee}{\end{equation}}
\newcommand{\ben}{\begin{displaymath}}
\newcommand{\een}{\end{displaymath}}
\newcommand{\ba}{\begin{eqnarray}}
\newcommand{\ea}{\end{eqnarray}}
\newcommand{\ban}{\begin{eqnarray*}}
\newcommand{\ean}{\end{eqnarray*}}

\newcommand{\eps}{\varepsilon}

\newcommand{\g}{\gamma}
\newcommand{\kvet}{\bf k}

\begin{document}
\begin{titlepage}
\vspace{1cm}

\begin{center}
{\huge \bf Electroweak Sudakov form factors and }
\\ \vspace{0.5cm}
{\huge \bf nonfactorizable soft QED effects}
\\ \vspace{0.5cm}
{\huge \bf at NLC energies}

\vspace{2cm}
{\Large
{\bf P. Ciafaloni$^{(a)}$ and  D. Comelli$^{(b)}$}}\\
\vspace{1cm}
{\it \large
(a) Dipartimento di Fisica \& INFN, via Arnesano, 73100 Lecce \\
(b) INFN sezione di Ferrara, via Paradiso 12, 44100 
Ferrara}\\

\vspace{5cm}
{\large\bf Abstract}
\end{center}
\begin{quotation}
We study the leading log infrared behavior of electroweak 
corrections at TeV scale energies, that will be reached by
next generation of linear colliders (NLC).
We show  that, contrary to what happens 
at typical LEP energies, it is not anymore possible to disentangle 
``pure electroweak'' 
from  ``photonic'' corrections. This means that soft QED effects
do not factorize and therefore
cannot be  treated in the usual ``naive'' way 
they were accounted for in the  LEP-era.
The nonfactorizable effects come up first at the two loop LL level, that we
calculate explicitly for a fermion source that is neutral under the
SU(2)$\otimes$U(1) gauge group (explicitly, a Z' decay into two fermions). 
The basic formalism we set up can be used to calculate LL 
effects at any order of perturbation theory. The results of this paper might
be important for future calculations of electroweak corrections
at NLC energies.
\end{quotation}
\vspace{3cm}
\end{titlepage}

\def\baselinestretch{1.1}

\section{Introduction}

With next generation of linear colliders (NLC \cite{NLC}), a new era in the
testing of electroweak interactions at the quantum level will begin. In fact,
when the c.m. energy is much higher than the electroweak scale of about 100
GeV, the pattern of radiative corrections is rather different from what
happens at LEP. It turns out that the high energy asymptotic behavior
is dictated by the infrared (IR) structure of the theory \cite{ciafacome1}, 
so that the leading
terms grow with the c.m. energy $\sqrt{s}$ 
like $g^2 \log^2\frac{\sqrt{s}}{M}$,
where $g$ is the gauge coupling and $M_W\approx M_Z\equiv M$ is
the weak scale. In refs. \cite{ciafacome1} and \cite{Beccaria:1999xd} the one
loop leading $\propto g^2 \log^2\frac{\sqrt{s}}{M}$ and 
subleading $\propto g^2 \log\frac{\sqrt{s}}{M}$ corrections were
calculated, while in  \cite{Kuhn:1999de}
an expression for the leading terms at all orders
in perturbation theory 
($\propto (g^2\log^2\frac{\sqrt{s}}{M})^n$) has been found.
A common feature of the cited works is that they all consider only ``pure
e.w.'' corrections to two fermion production, 
meaning that the photon contribution is not taken into
account.  However, it is not clear what is the interplay between photon and
W,Z bosons contributions at such high energies.
For  LEP 1-2 \cite{Altarelli:1989hv}, the leading infrared photonic (QED) 
corrections are taken into account separately 
and factorize with respect to the ``hard'' corrections, including the
mentioned ``pure
e.w.'' corrections.
One is tempted to extrapolate this approach to higher energies, using what we
call a ``naive'' approach, which would give the following for leading logs 
and in the limit of massless fermions:
\begin{itemize}
\item
virtual leading  IR
QED corrections factorize and exponentiate giving a factor
depending  on the ``photon mass'' $\lambda$, that acts as a IR cutoff.
\item
soft photon emission also factorizes, with a factor
depending on a typical 
experimental resolution $\Delta E$. 
\item
``pure ew'' virtual
corrections can be taken into account separately; at the LL level
their effects do
not exponentiate trivially but still factorize with respect to the Born level 
\cite{Kuhn:1999de}.
\end{itemize}
In this paper we show that this ``naive'' approach that is correct for LEP
energies, is no longer correct when the energies are much higher than the
e.w. scale. In the latter case, it is not anymore possible to separate ``pure
e.w.'' corrections from photonic corrections.
A correct treatment of radiative electroweak corrections is the following:
\begin{itemize}
\item
{\sl Complete} electroweak virtual corrections are calculated,
taking into account also the photon contribution. The photon is given a mass
$\lambda$ to regularize IR divergences.
\item
soft photon emission is calculated, with photons having energy less than
the experimental resolution ${\Delta E}$
\end{itemize}
In the limit of massless fermions we consider here, the effect
of soft  photon emission is basically that of substituting the photon
mass $\lambda$ 
with the experimental resolution $\Delta E$. 
Therefore we regularize the IR divergences of the virtual e.w.
corrections by giving the photon a mass $\lambda$, which has the physical
meaning of an experimental resolution on both energies and angles (of order 
$\frac{\lambda}{\sqrt{s}}$ for the latter).
We are thinking about experimental resolutions of the order of 
$\lambda\approx$ 10
GeV, much lower than the W and Z bosons mass so that a process 
with W- or Z- bremsstrahlung is experimentally resolved.
Of course, a realistic calculation taking into account all mass scales and
experimental cutoffs,
would be much more complicated and goes beyond the scope
of this work. 

The behavior of electroweak radiative
corrections at energies much higher than the
electroweak scale is very interesting from a theoretical point of view, since
the IR structure of a (spontaneously) broken gauge theory with mixing between
the gauge groups has never been
considered in the literature.
Moreover,
from a phenomenological point of view, 
the planned new generation of linear colliders with TeV scale
c.m. energy and very high luminosity \cite{NLC} should be able to test
experimentally such structure.

Bearing in mind simplicity, 
we consider the LL electroweak corrections to 
two fermion production by a vector boson  behaving like a
singlet with respect to the SU(2)$\otimes$U(1) group. 
To fix ideas, and in order to take a case with 
phenomenological interest, we study the two fermions decay rate of a
massive (mass $>$1 TeV) Z' gauge boson unmixed with the usual Z  boson
and belonging to a group
which commutes with the SM group.
We therefore consider this case as the simplest probe of the infrared
structure of electroweak corrections, having in addition phenomenological
interest. Moreover, the
basic formalism we set up and the general considerations we make are
relevant for a more general class of processes of interest at NLC energies.  

\section{Leading IR electroweak form factor}

The tree level
amplitude for Z' decay into a $f-\bar{f}$ couple of
massless fermions is given by:
\be
M_0\equiv M_0^L+M_0^R=
g^{Z'}_L\bar{u}(p_f)\gamma^\mu P_L v(p_{\bar{f}})\eps_\mu^*(q)+
g^{Z'}_R\bar{u}(p_f)\gamma^\mu P_R v(p_{\bar{f}})\eps_\mu^*(q)
\ee
where $\eps_\mu(q)$ is the polarization of the Z' with momentum
$q=p_f+p_{\bar{f}}$ and $p_f$ ($p_{\bar{f}}$) is the fermion (antifermion)
momentum. We identify $M_{Z'}^2\equiv {s}=2(p_f\cdot p_{\bar{f}})$ 
in the following. Since a difference in the
masses of Z and W bosons is negligible in LL approximation, we set
$M_Z\approx M_W=M=90$ GeV.
In the limit of massless fermions chirality is conserved, so radiative
corrections do not mix left and right fermions, that we can consider
separately. 

In order to compute the leading radiative corrections in the infrared region
$\sqrt{s}\gg w\gg M$, where $w$ is the virtual boson energy and $M$ its mass,
we use the method of soft insertions formulae, which are
widely used in QED \cite{low} and are known to provide 
in QCD \cite{cornwall,BCM}
the leading IR singularities at double log level. This method consists
in factorizing the softest virtual momentum $k^\mu$ by computing 
external line insertions only, and in iterating this procedure by setting
$k=0$ in the left-over diagram.

Since Goldstone bosons (and ghosts) do not couple to the external massless
fermions, we are led to consider only nearly on-shell gauge bosons which are
emitted and reabsorbed by an external fermion leg.
A  gauge boson attaches to a fermion line with an
amplitude proportional to the eikonal current $ \frac{p_f^\mu}{k p_f}$. 
We work in Feynman gauge 
with massless fermions $p_f^2=p_{\bar{f}}^2=0$,
so that diagrams in which a boson is emitted and reabsorbed 
by the same (fermion or antifermion) line 
do not contribute in leading log approximation. 

\begin{figure}[htb]\setlength{\unitlength}{1cm}
\begin{picture}(12,6)
\put(3.5,5.5){(a)}\put(8.5,5.5){(b)}\put(13.7,5.5){(c)}
\put(3.2,4){$W,Z,\g$}
\put(3,0){$Z'$}
\put(2,0){\epsfig{file=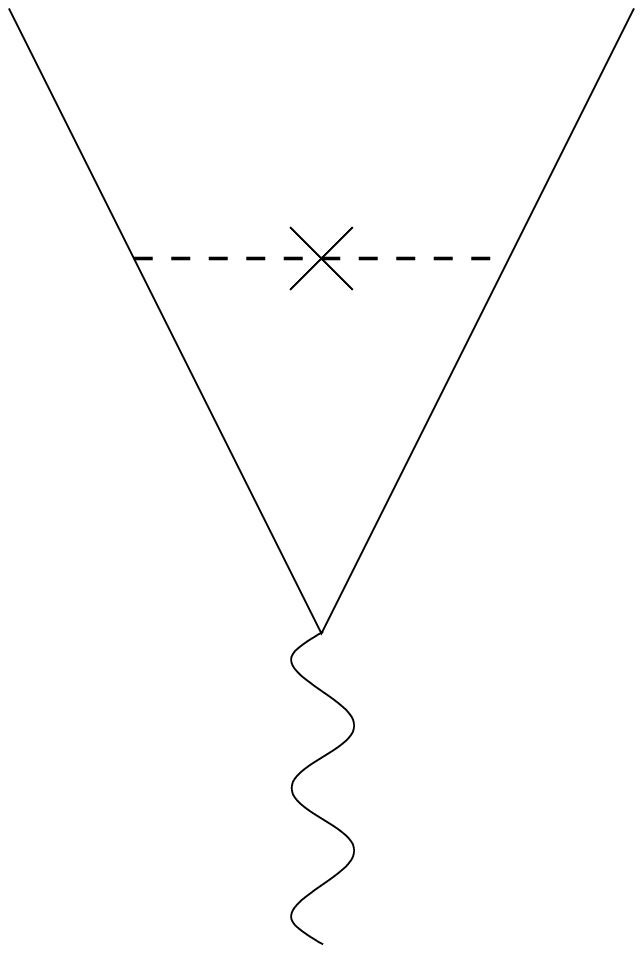,height=5cm}}
\put(6.5,4.5){$1$}
\put(7,3.5){$2$}
\put(8,2){$n$}
\put(8,0){$Z'$}
\put(7,0){\epsfig{file=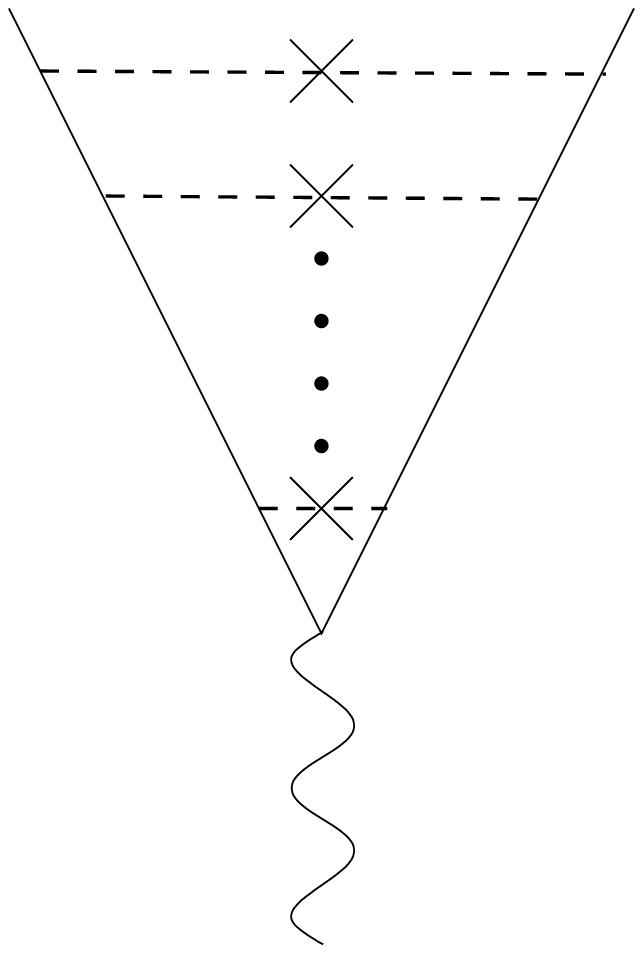,height=5cm}}
\put(12.4,4.5){QED}
\put(13.6,4.5){$\lambda<w<M$}
\put(12.4,3){EW}
\put(13.3,3){$M<w<\sqrt{s}$}
\put(12,0){\epsfig{file=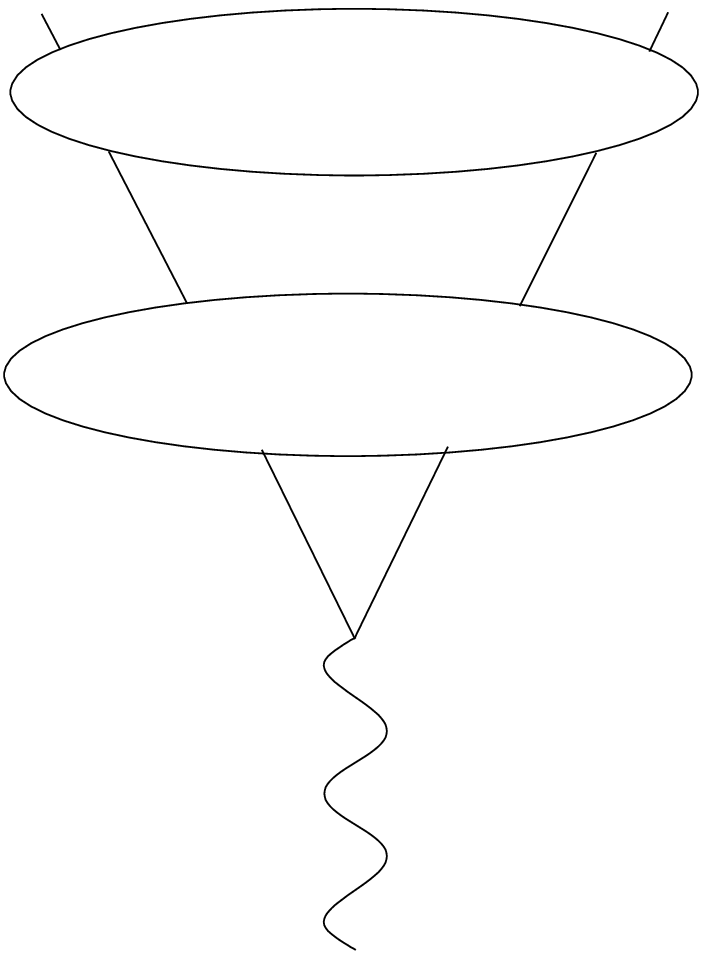,height=5cm}}
\end{picture}
\caption{(a)-(b)
:diagrams for soft boson insertion at 1 and n loops. Continuous lines
are fermion lines, and dashed lines are $W,Z,\g$ gauge bosons.Crosses
indicate that gauge bosons are close to mass-shell, and energies are such that
$w_1\ll w_2\ll.....\ll w_n$ (see text). (c): Pictorial representation of 
eq. (\ref{eqfig})}
\end{figure}
 
Following the calculations explicitly done in the Appendix,
we can now write the 1 loop correction to the tree level amplitude 
(see fig. 1a):
\be\label{1loop}
{\bf M}_1=-\sum_a\int d[k]\delta(k^2-M_a^2)
\bra f|J_\mu^a(k,p_f)J^{\mu a}(k,p_{\bar{f}})|f\ket M_0\qquad
J_\mu^a(k,p)=g^a\frac{p^\mu}{(kp)}T^a
\ee
where $|f\ket$ is a fermion belonging to a given representation of
the SU(2)$\otimes$U(1) gauge group and where we sum over all gauge
bosons $a$. The latter couple to fermions with SU(2)$\otimes$U(1) 
generators $T^a$ normalized to  to Tr$\{T^aT^b\}=\frac{1}{2}\delta^{ab}$. 
The charge operator is defined as $Q=T^3+Y$. We have defined
$d[k]=\pi\frac{d^4k}{(2\pi)^4}$; see Appendix for computational details. 
Notice that in general, the tree level amplitude
$M_0$ can have a complicated flavor structure 
\footnote{here and in the following,
by ``flavor'' we mean SU(2)$\otimes$U(1) quantum numbers; we only consider a
single generation of fermions}. 
On the other hand, since the Z' is a completely
neutral singlet under SU(2)$\otimes$U(1), the flavor structure is factorized
with respect to the tree level amplitude, as shown in (\ref{1loop}). This is
due to the fact that in this case $M_0$ is flavor diagonal and is therefore
just a number with respect to flavor. Notice
also that (\ref{1loop}) is just a shorthand notation; in fact 
the mass eigenstates ($Z,\g$) do not coincide with
the gauge bosons vectors $A_3,B$. Thus we have to take into 
account this mixing in the neutral sector (see later).

At higher orders, we assume that the soft boson insertion formula can be
iterated at the LL level in the ``strong energy ordering region''
\cite{BCM}. This means that the soft  bosons energies 
labelled from 1 (external boson) to $n$ (innermost boson) are such
that $w_1\ll w_2\ll.....\ll w_n$ (see fig. 1b).
Then, the two loop contribution is given simply by:
\be\label{M2}
{\bf M}_2=\sum_{a,b}\int d[k_1]d[k_2]\delta(k_1^2-M_a^2)\delta(k_2^2-M_b^2)
\bra f|J_\mu^a(k_1,p_f)J_\nu^b(k_2,p_f)J^{\nu b}(k_2,p_{\bar{f}})
J^{\mu a}(k_1,p_{\bar{f}})|f\ket M_0
\ee
and so on for the higher order contributions. In 
order to write the n-th order contribution  in a
compact form we make a slight change in the notation and we define 
the operators
$J^{\mu a}(k)=g^a \frac{p_f^\mu T^a}{k p_f}$,
$\tilde{J}^{\mu a}(k)=-g^a \frac{p_{\bar{f}}^\mu \tilde{T}^a}{k p_{\bar{f}}}$. 
The ``untilded'' operators act on the left as usual, i.e. 
$J^{\mu a}(k)\; M=g^a \frac{p_f^\mu}{k p_f}\;T^aM$, while the ``tilded''
operators
act on the right, i.e. $\tilde{J}^{\mu a}(k)\; M=g^a 
\frac{p_{\bar{f}}^\mu }{k p_{\bar{f}}}\;M{T}^a$.
Equation (\ref{M2}) can therefore be rewritten as:
\be
{\bf M}_2=\sum_{a,b}\int d[k_1]d[k_2]\delta(k_1^2-M_a^2)\delta(k_2^2-M_b^2)
\bra f|J_\mu^a(k_1)\tilde{J}^{\mu a}(k_1)
J_\nu^b(k_2)\tilde{J}^{\nu b}(k_2)
|f\ket M_0
\ee

In eikonal approximation, before integrating over $k_1,k_2.....k_n$,
the n-th order matrix element is obtained from
the matrix element of order n-1 by insertion of the following operator
$I(k)$:
\be
M_{n}= (-)^n\sum_a\delta (k^2-M_a^2)J^{\mu a}(k) 
M_{n-1} {J}^{\mu a}(k)
\equiv I(k)M_{n-1}
\qquad
I(k)=\sum_aJ^{\mu a}(k) \tilde{J}^{\mu a}(k)
\delta (k^2-M_a^2) 
\ee
Let us now calculate the insertion operator taking into account also gauge
bosons mixing in
the neutral sector. The insertion of a W boson gives 
$g^2(T^1\tilde{T}^1+T^2\tilde{T}^2)$ while inserting a Z boson gives
$\frac{g^2}{c_w^2}(T^3-s_w^2Q)(\tilde{T}^3-s_w^2\tilde{Q})
=g^2T^3\tilde{T}^3+g'^2Y\tilde{Y}-e^2Q\tilde{Q}$. By defining
$\bar{T}\cdot\bar{\tilde{T}}=\sum_iT_i\tilde{T}_i$ and since
we take the Z and W to be
degenerate at the same mass $M$, we can write:
\be\label{eq:3}
I(k)=\frac{p_fp_{\bar{f}}}{(p_fk)(p_{\bar{f}}k)}
\left\{
[g'^2Y\tilde{Y}+g^2\bar{T}\cdot\bar{\tilde{T}}-e^2Q\tilde{Q}]\delta(k^2-M^2)
+e^2Q\tilde{Q}\delta(k^2-\lambda^2)\right\}
\ee
The integrated matrix element is therefore:
\be\label{eq:4}
M_n =
\bra f|\int d[k_1]d[k_2]...d[k_n] I(k_1)I(k_2)...I(k_n)|f\ket
M_0\qquad w_1\ll w_2\ll....\ll w_n
\ee
For every  $k_i=(w_i,{\bf k}_i)$ we can now  integrate 
over the collinear regions ${\bf k}_i\cdot \hat{\bf {p}}_f\sim 1$,
${\bf k}_i\cdot \hat{\bf{p}}_{\bar{f}}\sim 1$ at fixed $w_i$. 
This gives factors $\log\frac{w_i}{M_a}$ (see Appendix).
The subsequent integral is over 
$\frac{dw_i}{w_i}=d\log w_i$. 
It is therefore natural to use the logarithm of the energy
as a variable, defining:
\be
l=\log\frac{M}{\lambda}\quad
L=\log\frac{\sqrt{s}}{M}\quad
x_i=\log\frac{w_i}{M}
\ee
The collinear integrals give then
$\log\frac{w}{\lambda}=x+l$ for the photon and 
$\log\frac{w}{M}=x$ for massive bosons. From eqns (\ref{eq:3},\ref{eq:4})
we obtain:

\be\label{Heff}
\int d[k]  I(k)=-\int dxH (x);\quad
H(x)=\Theta_0^L
[g'^2Y\tilde{Y}+g^2\bar{T}\cdot\bar{\tilde{T}}-e^2Q\tilde{Q}]x
+\Theta_{-l}^Le^2Q\tilde{Q}(x+l)\\
\ee
where $\Theta_a^b$ is 1 for $a<x<b$, 0 elsewhere.
Notice that here we have for sake of simplicity, reabsorbed all the constants
coming from phase space and integral over the angles in a redefinition of the
couplings; that is, we have made the substitution
$\frac{g_a^2}{2\pi^2}\to g_a^2$ (see Appendix).
We rewrite eq. (\ref{Heff}) in a slightly different form:
\ba
H(x)&=&\Theta_{-l}^0H_{QED}(x)+\Theta_0^LH_{EW}(w)\nonumber\\
H_{QED}(x)&=&e^2Q\tilde{Q}(x+l)\qquad
H_{EW}(x)=
(g'^2Y\tilde{Y}+g^2\bar{T}\cdot\bar{\tilde{T}})x
+e^2Q\tilde{Q}l\nonumber
\ea

Since the leading terms are given by the regions of phase space strongly
ordered in energy \cite{BCM}, the overall form factor  is given by:
\be
\sum_{n=0}^\infty {\bf M}_n=
\sum_{n=0}^\infty(-)^n\bra f|\int dx_1dx_2....
dx_n \underbrace{H(x_1)H(x_2)....H(x_n)}_{x_1<x_2<....<x_n}
 |f\ket M_0
\equiv\bra f | P_x \{\exp[-\int dxH(x) ]\} |f\ket M_0
\ee
The first thing to notice is that 
the total form factor can be  written 
\be\label{eqfig}
P_x\{\exp[-\int_{-l}^LdxH(x)]\}=P_x\{\exp[-\int_{-l}^0dxH_{QED}(x)]\}
\,\times\,
P_x\{\exp[-\int_{0}^LdxH_{EW}(x)]\}
\ee
Since $[H_{QED}(x_1),H_{QED}(x_2)]=0$, the
$x$-ordered exponential turns into a regular exponential for $-l<x<0$
and we have:
\be
\bra f|P_x\{\exp[\int_{-l}^0dx-H_{QED}(x)]\}
P_x\{\exp[-\int_{0}^LdxH_{EW}(x)]\}|f\ket
=\exp[-e^2q_f^2\frac{l^2}{2}]\bra f|P_x\exp[-\int_{0}^LdxH_{EW}(x)]|f\ket
\ee
We conclude that the total form factor is given by
\footnote{Operators are in capital letters, c-numbers in small
letters. Thus $Q$ is an operator with values $q_e=-1,q_\nu=0$}:
\be
\fbox{$
\sum_{n=0}^\infty {\bf M}_n=\exp[-e^2q_f^2\frac{l^2}{2}]
\sum_{n=0}^\infty { M}_n=
\exp[-e^2q_f^2\frac{l^2}{2}]\bra f|P_x\exp[-\int_{0}^LdxH_{EW}(x)]|f\ket M_0
$}
\ee\be\label{imp2}
\fbox{$
H_{EW}(x)=
(g'^2Y\tilde{Y}+g^2\bar{T}\cdot\bar{\tilde{T}})x
+e^2Q\tilde{Q}l
$}\ee
These formulae set up the basic formalism allowing to calculate the effects we
are interested in.
Here a separation of scales is implicit, in such a way that
\begin{itemize}
\item
For energies $\lambda<\sqrt{s}<M$
only the photon propagates, i.e. $H(x)=H_{QED}(x)$. Therefore  QED effects
still factorize and exponentiate, but this is true  only up to
an energy $M$.
\item
The contribution of photons with energies higher than $M$
is taken into account in a
nontrivial way in $H_{EW}$. 
The fact that the Z and $\g$ bosons have completely different mass scales 
$M,\lambda$ produces
an incomplete cancellation of
collinear divergences in the logarithmic term
$l=\log\frac{w}{\lambda}-\log\frac{w}{M}$ in (\ref{imp2}). 
\end{itemize}
This situation is depicted in fig. 1c. All photons with energies
between $\lambda$ and $M$ stand in the external ``blob'' and form an abelian
factorized structure. In the internal blob instead all bosons $W,Z,\g$ 
with $M<w<\sqrt{s}$
propagate.

Let us now  concentrate on that part of the  
form factor given by $H_{EW}$.
At the  1 loop level we obtain:
\be\label{eq:M1L}
M_1=-\bra f|\int_0^Ldx[(g'^2Y^2+g^2\bar{T}^2)x+e^2Q^2l]|f\ket M_0=
-(a_f\frac{L^2}{2}+b_flL)M_0
\ee
where $a_f$ and $b_f$ are the c-numbers\footnote{
for the fundamental representation of fermions, 
$\bar{T}^2=\sum_iT_iT_i=\frac{3}{4}$}:
\be\label{eq:aandb}
a_f=\bra f|g^2\bar{T}^2+g'^2Y^2|f\ket\qquad 
b_f=\bra f|e^2Q^2|f\ket
\ee
At 2 loops the algebra gets a bit more complicated. We have to calculate:
\be
M_2=\bra f|P_x\int_0^LdxH_{EW}(x_1)H_{EW}(x_2)|f\ket M_0
=\bra f|\int_0^Ldx_2\int_0^{x_2}dx_1H_{EW}(x_1)H_{EW}(x_2)|f\ket M_0
\ee
The operator with the highest value of $x$
in the $x$-ordered product, $H_{EW}(x_2)$ in this case,
corresponds to the more internal boson in fig.~1. Therefore, for this operator
we can substitute $\tilde{T}^a\to T^a\forall a$. 
This is of course not true for $H_{EW}(x_1)$. We have:
\ba\label{algebranonabeliana}
 H_{EW}(x_1)H_{EW}(x_2)=
H_{EW}(x_1)[(g'^2Y^2+g^2\bar{T}^2)x_2+e^2Q^2l]
\nonumber \\
=[(g'^2Y^2+g^2\bar{T}^2)x_2+e^2Q^2l][g'^2Y^2x_1+e^2Q^2l]
+\sum_{a=1}^3g^2x_1T^a[(g'^2Y^2+g^2\bar{T}^2)x_2+e^2Q^2l]T^a\\
=[(g'^2Y^2+g^2\bar{T}^2)x_2+e^2Q^2l][(g'^2Y^2+g^2\bar{T}^2)x_1+e^2Q^2l]
-2e^2g^2lYx_1T^3\nonumber 
\ea
where we have used the commutator algebra:
\ba
[((g'^2Y^2+g^2\bar{T}^2)x_2+e^2Q^2l),Q]
=[((g'^2Y^2+g^2\bar{T}^2)x_2+e^2Q^2l),Y]=0\\
\left[T^a,(g'^2Y^2+g^2\bar{T}^2)x_2+e^2Q^2l\right]=
e^2l [T^a,Q^2]=-2e^2lYi\eps^{3ab}T^b
\ea
Using (\ref{algebranonabeliana}) we easily obtain:
\be\label{eq:III}
M_2=
\left\{\bra f|P_x \int_0^Ldx_1dx_2H_{EW}(x_1)H_{EW}(x_2)|f\ket\right\}M_0
=
\left\{\frac{1}{2}(a_f\frac{L^2}{2}+b_flL)^2
-\frac{1}{3}e^2g^2lL^3y_ft^3_f\right\}M_0
\ee

It should be evident from the above that there is in general no simple
all-orders 
exponentiation of the LL effects. This is due to the combined effect of the
nonabelian group structure and of the gauge bosons mixing in the neutral
sector. Moreover, what we see is that there is an interesting interplay
between the $Z$ and $\g$ contributions. In particular, the term in
(\ref{imp2}) proportional to $Q\tilde{Q}$ is nonzero due to a non complete
cancellation of collinear divergences cutoffed by two different mass scales $M$
and $\lambda$.
\section{Comparing with the ``naive'' approach}
We now wish to compare our approach with
the ``naive'' approach where QED corrections are taken into account 
separately. They 
are supposed to factorize and exponentiate with a factor \cite{low}
$\exp[-e^2q_f^2\log^2\frac{\sqrt{s}}{\lambda}]=
\exp[-e^2q_f^2(l+L)^2]$.
We concentrate therefore on the ``pure electroweak'' part of the corrections.
These include only Z and W contributions, so that:
\ben
\Mnaif_n=
\bra f|\int d[k_1]d[k_2]...d[k_n] \Inaif (k_1)\Inaif (k_2)...\Inaif 
(k_n)|f\ket
\quad
\Inaif (k)=\frac{p_fp_{\bar{f}}}{(p_fk)(p_{\bar{f}}k)}
[g'^2Y\tilde{Y}+g^2\bar{T}\cdot\bar{\tilde{T}}-e^2Q\tilde{Q}]
\delta(k^2-M^2)
\een

\be\label{Heffnaif}
\int d[k] \Inaif (k)=-\int dx\Hnaif (x);\quad
\Hnaif(x)=
[g'^2Y\tilde{Y}+g^2\bar{T}\cdot\bar{\tilde{T}}-e^2Q\tilde{Q}]x
\ee
Since $[\Hnaif(x_1),\Hnaif(x_2)]=0$, $\int_0^L \Hnaif(x) dx$ exponentiates as
an operator, and the ``pure e.w.'' form factor is given by:
\be\label{eq:8}
\Mnaif=
\bra f|\exp\left\{-
[g'^2Y\tilde{Y}+g^2\bar{T}\cdot\bar{\tilde{T}}-e^2Q\tilde{Q}]\frac{L^2}{2}
\right\}
|f\ket M_0
\ee
We would like to stress the fact that 
the form factor exponentiates {\sl as an operator} 
and not as a number. Again, the
reason of non exponentiation are  noncommutativity and mixing. However,
(\ref{Heffnaif}) misses the presence of the scale $\lambda$ 
which is present in (\ref{imp2}).
By using commutator algebra, we obtain in this case the following results at 1
and 2 loop level:
\be\label{eq:MnaifL}
\Mnaif_1=
\bra f|-[g'^2Y^2+g^2\bar{T}^2-e^2Q^2]\frac{L^2}{2}|f\ket
 M_0=-(a_f-b_f)\frac{L^2}{2}M_0
\ee
\be\label{nonabelianonaif}
\Mnaif_2=\frac{1}{2}
\bra f|[(g'^2Y\tilde{Y}+g^2\bar{T}\cdot\bar{\tilde{T}}-e^2Q\tilde{Q})
\frac{L^2}{2}]^2|f\ket
M_0=\frac{1}{2}(\frac{L^2}{2})^2
[(a_f-b_f)^2+2e^2g^2y_ft^3_f]M_0
\ee
where $a_f$ and $b_f$ are defined in (\ref{eq:aandb}).

To compare the results of the previous sections, we have to include properly
QED effects that factorize in a different way in the two different approaches;
thus we define:
\be\label{eq:Delta}
\Delta\equiv
\exp[-e^2q_f^2\frac{l^2}{2}](\sum_{i=0}^{\infty}M_n)
-
\exp[-e^2q_f^2\frac{(l+L)^2}{2}](\sum_{i=0}^{\infty}\Mnaif_n)
=
\sum_{i=1}^{\infty}\Delta_i\qquad
\Mnaif_0=M_0
\ee
with the self-explaining convention that $\Delta_i$ is the difference 
of order $(g^2)^i$ between
the ``naive'' and the ``correct'' approach. From  
(\ref{eq:M1L},\ref{eq:MnaifL},\ref{eq:Delta})
we see that $\Delta_1=0$ both for left and right fermions. 
This was to be expected since at the
one loop level $\g$, Z, W contributions can all be treated separately.
At higher orders, let us now
consider separately left and right final fermions.

In the right fermions case, 
everything is proportional to the charge operator $Q$ and
therefore the operator algebra is commuting; we call this the abelian case.
It is easy  to show that for right fermions $\Delta=0$
to all orders.
In fact in this case the abelian structure allows for 
exponentiation of the one loop result, giving:
\be 
\exp[-e^2q_f^2\frac{l^2}{2}]\exp[-(g'^2q_f^2\frac{L^2}{2}+e^2q_f^2lL)]
M_0^R
=
\exp[-e^2q_f^2\frac{(l+L)^2}{2}]\exp[-(g'^2q_f^2-e^2q_f^2)\frac{L^2}{2}]
M_0^R
\ee
i.e., the same result with the two different approaches.

On the other hand, in the case of left handed fermions 
the second order value of $\Delta$ 
is different from zero: 
\be\label{deltaIII}
\Delta_2=-e^2g^2y_ft^3_f(\frac{L^4}{4}+\frac{L^3l}{3})
\ee
Several comments are in order. In first place, we have seen
that this expression arises
 mathematically from the fact
that the operators $Q$ and
 $T^1,T^2$ do not commute. Physically 
this corresponds to the fact  that the insertions
of a  W and of a photon
do not  commute due to the change of
flavor. 
The second observation is that  $\Delta_2$
contains a term proportional to $L^4$.
Now, the limit
$\frac{l}{L}\to 0$ (or, equivalently, $\sqrt{s}\to\infty$)
can be physically interpreted
 as the unbroken symmetry
limit, in which  the whole group SU(2)$\otimes$U(1) exponentiates.
However, we see from eq (\ref{deltaIII})
that $\Delta_2$  doesn't vanish in this limit; 
something must be wrong with one of the two approaches.
 The all orders formula 
in our approach can be easily evaluated in this limit
\be
\exp[-q_f^2\frac{l^2}{2}]\bra f|P_x\exp\int_{0}^LdxH_{EW}(x)|f\ket
M_0^{L}\stackrel{l\to 0}{\longrightarrow}
\bra f|P_x\exp\int_{0}^Ldx
[g'^2Y\tilde{Y}+g^2\bar{T}\cdot\bar{\tilde{T}}]x|f\ket M_0^{L}
\ee
and gives  the exponential of the SU(2)$\otimes$U(1)
group Casimir, 
$\exp[-(g'^2y_f^2+\frac{3}{4}g^2)\log^2\frac{\sqrt{s}}{M}]$.
On the contrary, from formula
 (\ref{deltaIII}), 
we see that the ``naive''
 approach gets a different result already
 at the 2 loop level.

We can summarize the situation  like this:
\begin{itemize}
\item
In the abelian case, the form factor is a regular exponential
and we get the same results with the two approaches.
\item
in the non abelian case, beginning from the 2 loop level
there is a difference due to commutator algebra.
\item
The ``naive'' approach doesn't get the right result 
in the limit $\sqrt{s}>>> M$ where we expect 
that SU(2)$\otimes$U(1)
factorizes as properly accounted for by  our approach.
\end{itemize}

To conclude, we observe that the numerical value of $\Delta$ depends on the
value of $\lambda$. However, for typical values
$\sqrt{s}\sim $ 1 TeV, $\lambda\sim$ 10 GeV,  
the difference between the two approaches is
$\Delta_2\approx 2\times 10^{-3}$
\footnote{we remind the reader that couplings have been
renormalized, so one should substitute 
$g_a^2\to\frac{g_a^2}{2\pi^2}$ in formulae
like (\ref{deltaIII}) to get the
results with the usual definitions of couplings}
and is therefore non negligible
in view of the expected experimental accuracy of a few {\it permille}
at NLCs.
\section{Conclusions}

We have investigated the leading IR
behavior of radiative electroweak corrections at energies much higher than the
electroweak scale.
The ``naive'' expectation of a pattern similar to the LEP case, with QED soft
effects factorizing in an independent way from the rest of ``pure e.w.''
corrections, turns out to be incomplete. Instead, photonic virtual
effects only exponentiate up to a scale $M$, which is of the order of the W
and Z bosons mass. Above $M$, the effects of
mixing  in the neutral sector and the presence of two mass scales
for the photon on one side and $W,Z$ bosons on the other 
have to be taken into account properly. We have established a formalism 
allowing to compute the LL electroweak corrections at any order, and 
calculated explicitly the second order effect in the case under exam.
We have seen two different reasons for the difference with the ``naive''
approach: one is the noncommutativity of SU(2)$\otimes$U(1) generators and 
the other is spontaneous symmetry breaking,
which induces mixing in the neutral sector and generates  
two different mass scales for the gauge bosons.

In this paper we have considered the
simple case of a process with 
two fermions on the external legs
coupling to a  vector boson
which is neutral under the SU(2)$\otimes$U(1) gauge group.
Cases of more immediate phenomenological interest, like 
processes with 4 fermions on the external legs
(for instance, $e^+e^-\to\mu^+\mu^-$), that 
have a more complicated flavor structure and a 
higher number of diagrams,
are currently under study.
However, the
basic formalism we have set up and the general considerations we have made
are 
relevant for a large class of processes of interest at NLC energies.

In general, for processes of relevance for NLCs  we
expect a difference between our approach and the ``naive'' one 
at the 2 loop LL level, of the order 
$(\frac{g^2}{16 \pi^2}\log^2\frac{s}{M^2})^2\sim$ 
a few {\sl permille}. Since this difference arises at the leading log level,
any future
calculation of higher order leading IR electroweak corrections 
at TeV scale energies
will have to cope with spontaneous symmetry breaking, mixing, 
and with the presence of two gauge bosons scales.
\vspace{1cm}

\noindent {\Large\bf Acknowledgments}

The authors are indebted to M. Ciafaloni for many useful discussions and
careful reading of the manuscript.

\vspace{1.cm}
\appendix\section{Appendix}
Let us consider the emission of a boson with momentum 
 $k=(w,{\bf k}),\,k^2=M^2$ from a fermion 
with momentum $p_f=\frac{\sqrt{s}}{2}(1,\hat{\bf p}_f),\,p_f^2=0$.
The denominator of the fermion propagator after the emission is given by:
\be\label{1bosone}
(p_f+k)^2=M^2+w\sqrt{s}
\left(1-\frac{\sqrt{w^2-M^2}}{w}\cos\theta\right)
\approx \frac{w\sqrt{s}}{2}\left(\theta^2+\frac{M^2}{w^2}
+\frac{2M^2}{w\sqrt{s}}\right)
\approx \frac{w\sqrt{s}}{2}\left(\theta^2+\frac{M^2}{w^2}\right)
\ee
where $\cos\theta=\hat{\bf k}\hat{\bf p}_f$ and where  we use the fact that 
the leading logs come from the region
$M\ll w\ll \sqrt{s},\theta\approx 0$. 
The collinear (lower) cutoff is therefore simply given 
by $\theta^2>\frac{M^2}{w^2}$. 
The integral relevant for the one loop result, whose leading (double log)
behavior is dictated by the
regions  of $k$ where the boson is on shell, is given by
\ba
i\frac{g^2}{(2\pi)^4}\int \frac{d^4k}{(k^2-M^2+i\eps)}
\frac{4(p_f\cdot p_{\bar{f}})}{(k^2+2p_f\cdot k)}
\frac{1}{(k^2+2p_{\bar{f}}\cdot k)}
\approx \frac{g^2}{(2\pi)^4}\pi 
(p_f\cdot p_{\bar{f}})\int\frac{\delta(k^2-M^2)d^4k}
{(p_f\cdot k)(p_{\bar{f}}\cdot k)}
\\
=\frac{g^2}{(2\pi)^4}\pi (p_f\cdot p_{\bar{f}})
\int\left. \frac{d^3{\bf k}}{w}\frac{1}{(p_f\cdot k)}
\frac{1}{(p_{\bar{f}}\cdot k)}\right|_{w=\sqrt{|\kvet|^2+M^2}}
=\frac{g^2}{(2\pi)^4}\pi^2\int\frac{|\kvet|^2 d |\kvet|}{w^3}
\int\frac{d\theta^2}{(\theta^2+\frac{M^2}{w^2})}
\\
=-\frac{g^2}{16\pi^2} \int_{M}^{\sqrt{s}}
\frac{dw}{w}{2}\log\frac{w}{M}=
-\frac{g^2}{16\pi^2}\int_0^{\log\frac{\sqrt{s}}{M}} d (\log^2\frac{w}{M})
=-\frac{g^2}{16\pi^2}\log^2\frac{\sqrt{s}}{M}
\ea
The integral receives two equal contributions from the regions
where $\cos\theta=\hat{\bf k}\hat{\bf p}_f\approx 1$
and where $\cos\bar{\theta}=\hat{\bf k}\hat{\bf p}_{\bar f}\approx 1$
Moreover, since 
$\delta(k^2-M^2)$ has two poles, we have to consider also 
$w=-\sqrt{|\kvet|^2+M^2}$. This second pole gives the same results, but in
different angular regions ($\cos\theta\to-\cos\theta$). In the end we have
to multiply by a factor 4. The final result 
is $g^2/(4\pi^2)\log^2\frac{\sqrt{s}}{M}$. This result agrees with the one
obtained in \cite{ciafacome1} by calculating the C-functions asymptotic
behavior.
In eq.~(\ref{Heff}), we redefine
$g^2/(2\pi^2)\to g^2$ and this result is written 
$g^2/2\log^2\frac{\sqrt{s}}{M}$.

\end{document}